# Energy distribution and energy fluctuation in Tsallis statistics

Guo Ran,    Du Jiulin[*]

*Department of Physics, School of Science, Tianjin University, Tianjin 300072, China*

**Abstract:** The energy distribution and the energy fluctuation in the Tsallis canonical ensemble are studied with the OLM formalism but following a new way. The resulting formula for the energy fluctuation is not the same as that in previous work [Liu L.Y. and Du J.L., Physica A **387**(2008)5417]. In discussing the application of an ideal gas, we find that the energy fluctuation can not be negligible in the thermodynamic limit, showing the ensemble nonequivalence for this case in Tsallis statistics. We investigate the energy fluctuation with a Tsallis generalized canonical distribution studied by Plastino and Plastino [Phys. Lett. A **193**(1994)140] for describing a system in contact with a finite heath bath. For this situation, the two formulae for the energy fluctuation are shown to be equivalent, while the nonextensive parameter $q$ plays an very important role.

**PACS:** 05.20.-y; 05.20.Gg
**Key words:** Energy distribution; Energy fluctuation; Tsallis statistics.

---

[*] E-mail addresses: jiulindu@yahoo.com.cn

## 1. Introduction

Tsallis statistics is a logical generalization of Boltzmann-Gibbs (BG) statistics, which has attracted a great deal of attention, and it has made quite a lot of applications in investigating some interesting problems in the fields of including physics, astronomy, chemistry, biology, economics and engineering [1]. It is known as nonextensive statistical mechanics. As we know, the energy fluctuation of canonical ensemble is very important in statistical mechanics. In BG statistics, the energy fluctuation is very small so that it can be negligible when particle numbers of a system tend to infinity and thus there is no difference between the canonical ensemble and the microcanonical ensemble. However, for Tsallis statistics, the question is wether this characteristic in BG statistics can be still kept and further wether this characteristic depends on different ways for calculating the energy fluctuation. In thes investigations, what roles does the nonextensiv parameter $q$ play? In this work, following a new way we investigate the energy distribution, the energy fluctuation, the ensemble equivalence in Tsallis statistics and the association with the ways to calculate the energy fluctuation. Some important parameters that are related to these formulae, such as $q$, will be discussed.

The paper is organized as follows. In Sec.2, we introduce some of general formulae for the probability distribution functions in Tsallis statistics. In Sec.3, following a new way we study the energy distribution and the energy fluctuation. In Sec.4, we discuss the parameters, the energy fluctuation and the ensemble equivalence in Tsallis statistics. Finally in Sec.5, we give the conclusions.

## 2. The distribution functions in Tsallis statistics

Tsallis entropy [2] is given by

$$S_q \equiv k \frac{\sum_{i=1}^{w} p_i^q - 1}{1 - q},\tag{1}$$

where $k$ is Boltzmann constant, $p_i$ is the probability that the system under consideration is in its $i$th configuration, $W$ is the total numbers of the configurations, and $q$ is a nonextensive parameter whose deviation from unity measures the degree of nonextensivity of the system. BG entropy can be obtained from Tsallis entropy if we take the limit $q \to 1$.

Maximizing Tsallis entropy subject to the nomorlization condition and the energy constraint [3],

$$\sum_{i=1}^{W} p_i - 1 = 0 , \quad \text{and} \quad \sum_{i=1}^{W} p_i^q \left( E_i - U_q \right) = 0 , \tag{2}$$

one can derive the probability distribution function for the energy $E_i$,

$$p_i \left( E_i \right) = \frac{1}{Z_q} \left[ 1 - (1-q) \frac{\beta}{c_q} \left( E_i - U_q \right) \right]^{\frac{1}{1-q}} , \tag{3}$$

where $\beta$ is the Lagrange multiplier associated with the energy, and $Z_q$ is the partition function,

$$Z_q = \sum_{i=1}^{W} \left[ 1 - (1-q) \frac{\beta}{c_q} \left( E_i - U_q \right) \right]^{\frac{1}{1-q}} , \tag{4}$$

with

$$c_q = \sum_{i=1}^{W} p_i^q = 1 + \frac{1-q}{k} S_q \left( U_q \right) . \tag{5}$$

Usually, $\beta$ is identified with the inverse temperature, related [3] by $k\beta = \partial S_q / \partial U_q$. If one uses a new parameter $\beta_q$ to denote

$$\beta_q = \frac{\beta}{c_q} , \tag{6}$$

and call the parameter $T_q = 1/(k\beta_q)$ physical temperature [4], one can write the probability distribution function as

$$p_i \left( E_i \right) = \frac{1}{Z_q} \left[ 1 - (1-q) \beta_q \left( E_i - U_q \right) \right]^{\frac{1}{1-q}} , \tag{7}$$

and the partition function becomes

$$Z_q = \sum_{i=1}^{W} \left[ 1 - (1-q) \beta_q \left( E_i - U_q \right) \right]^{\frac{1}{1-q}} , \tag{8}$$

which are consistent with those in optimal Lagrange multipliers (OLM) formalism [5].

## 3. The energy distribution and the energy fluctuation

In BG statistics, there are two ways to calculate the energy fluctuation [6]. The first way is to use the formula

$$\langle E^2 \rangle - \langle E \rangle^2 = -\frac{\partial U}{\partial \beta} = kT^2 C_V , \qquad (9)$$

and then to obtain the relative fluctuation of energy by

$$\frac{\sqrt{\langle E^2 \rangle - \langle E \rangle^2}}{U} = \frac{\sqrt{kT^2 C_V}}{U} . \qquad (10)$$

This formula was used to study the energy fluctuation of canonical ensemble in Tsallis statistics [7]. The second way is to expand the function of energy distribution at a point of internal energy to derive the relative fluctuation, the result of which is the same as that by the first way. Here we follow the second way to calculate the relative fluctuaion of energy in Tsallis statistics.

In Tsallis statistics, the expectation value of one physical quantity $O$ can be defined by the so-called normalized $q$-expectation [3] , i.e.

$$\langle O \rangle_q = \frac{\sum_{i=1}^{W} p_i^q (E_i) O}{\sum_{i=1}^{W} p_i^q (E_i)} , \qquad (11)$$

with the state distribution function, $p_i(E_i)$, where $\sum_i$ is over all the micro-states with energy $E_i$. Eq.(11) can also be written for the energy distribution function, $p(E)$, as

$$\langle O \rangle_q = \frac{\sum_E p^q (E) O}{\sum_E p^q (E)} , \qquad (12)$$

where $\sum_E$ is over all the possible values of the energy. At each energy there may be many micro-states. Thus, the energy distribution function, $p(E_i)$, is

$$p(E_i) = \omega^{\frac{1}{q}} (E_i) p_i (E_i) , \qquad (13)$$

where $\omega(E_i)$ is the micro-state numbers of the system at the energy $E_i$. Substituting Eq.(7) into the right hand side of Eq.(13), one has

$$p(E_i) \sim \omega^{\frac{1}{q}} (E_i) \left[ 1 - (1-q) \beta_q (E_i - U_q) \right]^{\frac{1}{1-q}} \qquad (14)$$

Without loss of generality, $E_i$ can be replaced by $E$ in Eq.(14) and then the energy distribution function can be expressed as

$$p(E) \sim \exp_q \left[ \ln_q \omega^{\frac{1}{q}}(E) \right] \exp_q \left[ -\beta_q (E - U_q) \right],$$  (15)

where the q-logarithmic and the q-exponential functions are respectively

$$\ln_q x = \frac{x^{1-q} - 1}{1-q}, \quad \text{and} \quad \exp_q(x) = \left[ 1 + (1-q) x \right]^{\frac{1}{1-q}}.$$  (16)

If the abbreviation $R_q$ denotes

$$R_q(E) = k \ln_q \omega^{\frac{1}{q}}(E),$$  (17)

then Eq.(15) is written as

$$p(E) \sim \exp_q \left\{ \frac{R_q(E)}{k} - \beta_q (E - U_q) \left[ 1 + (1-q) \frac{R_q(E)}{k} \right] \right\},$$  (18)

and a relation between $R_q$ and the entropy $S_q$ is found by the Tsallis entropy with the form of $S_q(E) = k \ln_q \omega(E)$ to be

$$R_q(E) = \frac{k}{1-q} \left\{ \left( 1 + \frac{1-q}{k} S_q(E) \right)^{\frac{1}{q}} - 1 \right\}.$$  (19)

We let the function in Eq.(18) be $I(E)$, given by

$$I(E) = \frac{R_q(E)}{k} - \beta_q (E - U_q) \left[ 1 + (1-q) \frac{R_q(E)}{k} \right],$$  (20)

and then we get its derivative at $E = E_m$,

$$\left. \frac{\partial I}{\partial E} \right|_{E=E_m} = b_q(E_m) \left\{ \frac{1}{q} \tilde{\beta}_q(E_m) \left[ 1 - (1-q) \beta_q (E_m - U_q) \right] - \beta_q \right\},$$  (21)

with

$$b_q(E_m) = 1 + \frac{1-q}{k} R_q(E_m)$$  (22)

and, in terms of the relation Eq.(19),

$$\tilde{\beta}_q(E_m) = \frac{1}{kc_q} \left. \frac{\partial S_q(E)}{\partial E} \right|_{E=E_m},$$  (23)

If $E_m$ is the energy at which the function $I$ takes its maximum, there is $\left( \partial I / \partial E \right)_{E=E_m} = 0$

and because $\exp_q(x)$ is a monotonically increasing function for $x$, $\exp_q(I)$ also take its maximum at the energy $E_m$. Thus $E_m$ is the value at which the distribution function Eq.(18) may get its maximum. Now the function $I(E)$ can be expanded as a series about $E_m$. The first order term vanishes, and then,

$$I(E) = I(E_m) + \frac{1}{2} \frac{\partial^2 I}{\partial E^2}\bigg|_{E=E_m} (E-E_m)^2 + \cdots . \qquad (24)$$

Neglecting those terms higher than the second order, it becomes

$$I(E) = I(E_m) - \frac{1}{2} \tilde{\beta}_q(E_m) \beta_q b_q(E_m) \left( \frac{k}{\tilde{C}_V} + \frac{1-q}{q} \right) (E-E_m)^2 , \qquad (25)$$

where the quantity $\tilde{C}_V$ is

$$\tilde{C}_V = -k\tilde{\beta}_q^{\,2} \frac{\partial E}{\partial \tilde{\beta}_q}\bigg|_{E=E_m} . \qquad (26)$$

Substituting Eq.(25) into Eq.(18), i.e. $p(E) \sim \exp_q[I(E)]$, it is

$$p(E) \sim \left[ 1 - (1-q) \frac{\tilde{\beta}_q^{\,2}(E_m)}{2q} \left( \frac{k}{\tilde{C}_V} + \frac{1-q}{q} \right)(E-E_m)^2 \right]^{\frac{1}{1-q}} . \qquad (27)$$

Using Eq.(27), we can calculate the $q$-average of the energy (see Appendix A.): $\langle E \rangle_q = E_m$. Because in Tsallis statistics it has been defind $\langle E \rangle_q = U_q$, we get $E_m = U_q$. Using Eq.(21), $(\partial I/\partial E)_{E=E_m} = 0$, a relation between $\tilde{\beta}_q$ and $\beta_q$ is found to be

$$\tilde{\beta}_q(U_q) = q\beta_q . \qquad (28)$$

Consequently, the energy distribution function Eq.(27) becomes

$$p(E) \sim \left[ 1 - (1-q) \frac{\beta_q^{\,2}}{2} \left( \frac{k}{C_V} + 1 - q \right)(E-U_q)^2 \right]^{\frac{1}{1-q}} , \qquad (29)$$

whith $C_V = \tilde{C}_V / q$ and

$$C_V = -k\beta_q^{\,2} \frac{\partial E}{\partial \beta_q} . \qquad (30)$$

Obviously, as expected, the Boltzmann distribution $p(E) \sim \exp\left[ -E/(2kT^2 C_V) \right]$ can be obtained by Eq.(29) if we take $q=1$. Noting that $p(E)$ is an energy distribution and it should reach a maximum at $E=E_m$ ( and thus E=$U_q$), this means that the second derivative for $E$ of $p(E)$ at $E=E_m$ must be less than zero, i.e.

$$\left.\frac{\partial^2 p(E)}{\partial E^2}\right|_{E=U_q} = -\beta_q^{\ 2}\left(\frac{k}{C_V}+1-q\right) < 0 . \qquad (31)$$

Accordingly, we find

$$q < 1 + \frac{k}{C_V} . \qquad (32)$$

Now we can calculate the relative fluctuation of energy by using Eq.(29). The result (see Appendix B.) is

$$\frac{\sqrt{\left\langle \left(E-U_q\right)^2\right\rangle_q}}{U_q} = \frac{1}{\beta_q U_q}\sqrt{\frac{2}{3-q}\frac{C_V}{(1-q)C_V+k}}, \quad \text{for } q<3. \qquad (33)$$

The standard form of the energy fluctuation in Boltzmann statistics, $\sqrt{C_V/k}/(\beta U)$, can be recovered from Eq.(33) when onee takes $q=1$.

As an example, we take an ideal gas to analyze the relative fluctuation of energy. For a classical ideal gas, the internal energy in the Tsallis canonical ensemble [8] is $U_q=3N/2\beta_q$, the heat capacity is $C_V=3Nk/2$, and thus, from Eq.(33), the relative fluctuation of energy reads

$$\frac{\sqrt{\left\langle \left(E-U_q\right)^2\right\rangle_q}}{U_q} = \frac{2}{3N}\sqrt{\frac{2}{3-q}\frac{3N}{(1-q)3N+2}} . \qquad (34)$$

Because of $N(q-1)=\text{const.}$[9], it becomes

$$\frac{\sqrt{\left\langle \left(E-U_q\right)^2\right\rangle_q}}{U_q} \propto \sqrt{\frac{1}{2N-\text{const.}}} . \qquad (35)$$

It is clear that if there is $q \neq 1$, the particle numbers $N$ must be finite. Hence the relative fluctuation of energy also must be finite, which appears different from the case in BG statistics.

## 4. The parameters, the energy fluctuation and the ensemble equivalence

We consider that the Tsallis generalized canonical distribution describes a system in contact with a finite heath bath, studied in Ref. [10] by Plastino et al, where the Tsallis distribution is assumed to describe a system whose number of states $\eta(E)$ increases according to a power-law for the energy $E$ with an index $\alpha - 1$, i.e. $\eta(E) \propto E^{\alpha-1}$. The probability distribution reads

$$P_j = Z^{-1}\left(1 - \varepsilon_j/E_0\right)^{\alpha-1}, \tag{36}$$

where the normalization constant is $Z = \sum_j \left(1 - \varepsilon_j/E_0\right)^{\alpha-1}$. On the other hand, the escort probability distribution [11],

$$P_i = \frac{p_i^q(E_i)}{c_q} = \frac{\left[1 - (1-q)\beta_q\left(E_i - U_q\right)\right]^{\frac{q}{1-q}}}{\sum_{i=1}^{w}\left[1 - (1-q)\beta_q\left(E_i - U_q\right)\right]^{\frac{q}{1-q}}}, \tag{37}$$

may be considered as a real probability. Comparing Eq.(36) with Eq.(37), we can determine the relations between these important parameters $q$, $\alpha$, $\beta_q$, $U_q$ and $E_0$, i.e.

$$q = 1 - \frac{1}{\alpha} \quad \text{and} \quad \beta_q = \frac{\alpha}{E_0 - U_q}. \tag{38}$$

For $D$-dimension classical ideal gas, $\alpha = DN/2$, the heat capacity in canonial ensemble [8] is $C_V = DNk/2$. The relations between $C_V$ and $\alpha$ as well $q$ are shown as follows.

For the microcanonical ensemble, Tsallis entropy is written as

$$S_q = \frac{k}{q-1}\left(1 - \eta^{1-q}\right). \tag{39}$$

By using Eq.(6) and Eq.(39), we can write the inverse physical temperature as $T_q = E/k(\alpha-1)$. And then the heat capacity $\tilde{C}_V = \partial E/\partial T_q$ is $\tilde{C}_V = k(\alpha-1)$. According to Eq.(30), the heat capacity in canonical ensemble, $C_V = \tilde{C}_V/q$, is found to be $C_V = k\alpha$, and therefore the parameter $q$ is related to the heat capacity by

$$q = 1 - \frac{k}{C_V}. \tag{40}$$

The formula of relative fluctuation of energy [7], derived by Liu and Du following

the first way, is

$$\frac{\sqrt{\left\langle \left(E - U_q\right)^2\right\rangle_q}}{U_q} = \frac{1}{\beta_q U_q}\sqrt{\frac{C_V}{\left(1-q\right)C_V + \left(2-q\right)k}} \ . \tag{41}$$

Its form seems different from Eq.(33) at first sight. However, when substituting Eq.(40) for the $q$ parameter into Eq.(41) and Eq.(33), respectively, we find that these two formulae for the energy fluctustion are identical, which therefore suggests that the nonextensive parameter $q$ plays an important role in these investigations.

Eq.(40) suggests that if the Tsallis generalized canonical distribution can be employed to describe a system whose number of states is proportional to a power-law for the energy $E$ with an index $\alpha - 1$, the nonextensive parameter $q$ is related to the heat capactiy of the system. If the particle numbers in the system goes to infinite, the heat capactiy tends to infinite, which is the case if and only if $q \to 1$ on the basis of Eq.(40). Namely, the Tsallis canonical distribution can describe a system with finite particle numbers and in contact with a finite heat bath. This result shows the nonequivalence between the microcanonical ensemble and the canonical ensemble in Tsallis statistics.

## 5. Conclusions

We have studied the energy distribution and the energy fluctuation in the Tsallis canonical ensemble with the OLM formalism by employing a new way. The resulting formula for the energy fluctuation is presented generally by Eq.(33), the form of which is not the same as that in previous work by Liu and Du [7]. We consider a Tsallis generalized canonical distribution, studied in [10] by Plastino and Plastino for describing a system in contact with a finite heath bath, whose number of states is proportional to a power-law for the energy $E$ with an index $\alpha - 1$. By comparing these canonical distributions, we have obtained the relations given in Eq.(38) between some important parameters such as $q$, $\alpha$, $\beta_q$, $U_q$ and $E_0$. The nonextensive parameter $q$ is found to relate the heat capactiy of the system by Eq.(40). In this situation, we have shown these two formulae for the energy fluctustion are equivalent, where the nonextensive parameter $q$ plays an very important role, and we have shown the nonequivalence between the microcanonical ensemble and the canonical ensemble in Tsallis statistics.

**Acknowledgements**

This work is supported by the National Natural Science Foundation of China under Grant No.11175128.

**Appendix A.** Calculation of $\langle E \rangle_q$

**a.** For the case of $q > 1$:

$$\langle E \rangle_q = \frac{\int_0^{+\infty} \left[ p(E) \right]^q E \, dE}{\int_0^{+\infty} \left[ p(E) \right]^q \, dE}$$

$$= \frac{\int_0^{+\infty} \left\{ 1 - (1-q) \left[ \frac{\tilde{\beta}_q^{\,2}}{2q} \left( \frac{k}{\tilde{C}_V} + \frac{1-q}{q} \right) (E - E_m)^2 \right] \right\}^{\frac{q}{1-q}} E \, dE}{\int_0^{+\infty} \left\{ 1 - (1-q) \left[ \frac{\beta_q^{*2}}{2q} \left( \frac{k}{\tilde{C}_V} + \frac{1-q}{q} \right) (E - E_m)^2 \right] \right\}^{\frac{q}{1-q}} dE} . \tag{A.1}$$

Let $x = E - E_m$, it becomes

$$\langle E \rangle_q = \frac{\int_{-E_m}^{+\infty} \left\{ 1 - (1-q) \left[ \frac{\tilde{\beta}_q^{\,2}}{2q} \left( \frac{k}{\tilde{C}_V} + \frac{1-q}{q} \right) x^2 \right] \right\}^{\frac{q}{1-q}} (x + E_m) \, dx}{\int_{-E_m}^{+\infty} \left\{ 1 - (1-q) \left[ \frac{\tilde{\beta}_q^{\,2}}{2q} \left( \frac{k}{\tilde{C}_V} + \frac{1-q}{q} \right) x^2 \right] \right\}^{\frac{q}{1-q}} dx} . \tag{A.2}$$

Since there should be $E_m = U_q + \text{const.}$, and generally $U_q$ is proportional to particle number $N$ of the system, and $N$ can be infinity for a macroscopic system, one has

$$\langle E \rangle_q \simeq \frac{\int_{-\infty}^{+\infty} \left\{ 1 - (1-q) \left[ \frac{\tilde{\beta}_q^{\,2}}{2q} \left( \frac{k}{\tilde{C}_V} + \frac{1-q}{q} \right) x^2 \right] \right\}^{\frac{q}{1-q}} (x + E_m) \, dx}{\int_{-\infty}^{+\infty} \left\{ 1 - (1-q) \left[ \frac{\tilde{\beta}_q^{\,2}}{2q} \left( \frac{k}{\tilde{C}_V} + \frac{1-q}{q} \right) x^2 \right] \right\}^{\frac{q}{1-q}} dx} = E_m . \tag{A.3}$$

**b.** For the case of $q < 1$:

In the same way as above, if let $x = E - E_m$, the $q$-average of energy is

$$\langle E \rangle_q = \frac{\int_{-a}^{a} \left\{ 1 - (1-q) \left[ \frac{\tilde{\beta}_q^{\,2}}{2q} \left( \frac{k}{\tilde{C}_V} + \frac{1-q}{q} \right) x^2 \right] \right\}^{\frac{q}{1-q}} (x + E_m)\, dx}{\int_{-a}^{a} \left\{ 1 - (1-q) \left[ \frac{\tilde{\beta}_q^{\,2}}{2q} \left( \frac{k}{\tilde{C}_V} + \frac{1-q}{q} \right) x^2 \right] \right\}^{\frac{q}{1-q}} dx}, \qquad \text{(A.4)}$$

with

$$a = \sqrt{\frac{2q^2}{(1-q)\tilde{\beta}_q^{\,2}} \frac{\tilde{C}_V}{(1-q)\tilde{C}_V + qk}}, \qquad \text{(A.5)}$$

where the cut-off condition of $q$-exponential function has been used, i.e.

$$\exp_q(x) = \begin{cases} \left[ 1 + (1-q)x \right]^{1/1-q}, & \text{for } 1 + (1-q)x > 0 \\ 0, & \text{otherwise} \end{cases}. \qquad \text{(A.6)}$$

Thus one obtains the $q$-average value of energy,

$$\langle E \rangle_q = E_m. \qquad \text{(A.7)}$$

**Appendix B.** Calculation of $\left\langle \left( E - U_q \right)^2 \right\rangle_q$

**a.** The case of $q > 1$:

$$\left\langle \left( E - U_q \right)^2 \right\rangle_q = \frac{\int_0^{+\infty} \left\{ 1 - (1-q) \left[ \frac{\beta_q^{\,2}}{2} \left( \frac{k}{C_V} + 1 - q \right) \left( E - U_q \right)^2 \right] \right\}^{\frac{q}{1-q}} \left( E - U_q \right)^2 dE}{\int_0^{+\infty} \left\{ 1 - (1-q) \left[ \frac{\beta_q^{\,2}}{2} \left( \frac{k}{C_V} + 1 - q \right) \left( E - U_q \right)^2 \right] \right\}^{\frac{q}{1-q}} dE}. \qquad \text{(B.1)}$$

Let $x = E - U_q$ and

$$\gamma = -\frac{(1-q)\beta_q^{\,2}}{2} \left( \frac{k}{C_V} + 1 - q \right), \qquad \text{(B.2)}$$

one derive, only if $1 < q < 3$ (if this condition is not satisfied, the integral will be divergent),

$$\left\langle \left( E - U_q \right)^2 \right\rangle_q \simeq \frac{\int_{-\infty}^{+\infty} \left\{ 1 + \gamma x^2 \right\}^{\frac{q}{1-q}} x^2 dx}{\int_{-\infty}^{+\infty} \left\{ 1 + \gamma x^2 \right\}^{\frac{q}{1-q}} dx} = \frac{\int_0^{+\infty} \left\{ 1 + \gamma x^2 \right\}^{\frac{q}{1-q}} x^2 dx}{\int_0^{+\infty} \left\{ 1 + \gamma x^2 \right\}^{\frac{q}{1-q}} dx}$$

$$= \frac{1}{\gamma} \frac{B\left(\frac{3}{2}, -\frac{3}{2} + \frac{q}{q-1}\right)}{B\left(\frac{1}{2}, -\frac{1}{2} + \frac{q}{q-1}\right)} = \frac{q-1}{\gamma(3-q)}, \tag{B.3}$$

where the Beta function is

$$B(p,q) = \int_0^{+\infty} \frac{y^{p-1}}{(1+y)^{p+q}} dy = \frac{\Gamma(p)\Gamma(q)}{\Gamma(p+q)}. \tag{B.4}$$

Thus one finds

$$\left\langle \left(E - U_q\right)^2 \right\rangle_q = \frac{2}{(3-q)\beta_q^2} \frac{C_V}{k + (1-q)C_V}, \quad \text{for } 1 < q < 3. \tag{B.5}$$

**b.** The case of $q < 1$:

Let $x = E - U_q$ and

$$\gamma' = \frac{(1-q)\beta_q^2}{2}\left(\frac{k}{C_V} + 1 - q\right), \tag{B.6}$$

one has

$$\left\langle \left(E - U_q\right)^2 \right\rangle_q = \frac{\int_{-a}^{a}\left\{1 - \gamma'x^2\right\}^{\frac{q}{1-q}} x^2 dx}{\int_{-a}^{a}\left\{1 - \gamma'x^2\right\}^{\frac{q}{1-q}} dx} = \frac{1}{\gamma'} \frac{B\left(\frac{3}{2}, 1 + \frac{q}{1-q}\right)}{B\left(\frac{1}{2}, 1 + \frac{q}{1-q}\right)}$$

$$= \frac{1-q}{\gamma'(3-q)} = \frac{2}{(3-q)\beta_q^2} \frac{C_V}{k + (1-q)C_V}. \tag{B.7}$$